\documentclass[aps,prd,onecolumn,nofootinbib,superscriptaddress,longbibliography]{revtex4-2}

\usepackage[utf8]{inputenc}
\usepackage[T1]{fontenc}
\usepackage{amsmath,amssymb,amsfonts,bm,mathtools}
\usepackage{physics}
\usepackage{graphicx}
\graphicspath{{prd_figures_v3/}}
\usepackage{xcolor}
\usepackage{hyperref}
\usepackage{booktabs}
\usepackage{microtype}
\usepackage{enumitem}
\usepackage{array}
\usepackage{tikz}
\usetikzlibrary{arrows.meta,positioning,shapes.geometric}

\hypersetup{colorlinks=true,linkcolor=blue,citecolor=blue,urlcolor=blue}

\renewcommand{\dd}{\mathrm{d}}
\newcommand{\cL}{\mathcal{L}}

\newcommand{\cD}{\mathcal{D}}
\newcommand{\cH}{\mathcal{H}}
\newcommand{\cO}{\mathcal{O}}

\renewcommand{\sech}{\operatorname{sech}}

\begin{document}

\title{Solitonic Construction of Artificial Neural Networks from Nonlinear Field Theory}

\author{R. A. C. Correa}
\email{rafael.a.correa@capgemini.com}
\affiliation{Capgemini, S\~ao Paulo Corporate Towers, Av. Pres. Juscelino Kubitschek 1909, S\~ao Paulo, SP 04543-907, Brazil}

\date{\today}

\begin{abstract}
We present a field-theoretic construction of a class of artificial neural networks from solitonic
degrees of freedom in nonlinear scalar field theory. The purpose is not to rename a standard
neural layer in the language of solitons, but to start from a continuum action, restrict the theory
to a nonperturbative sector containing localized stable solutions, perform a collective-coordinate
reduction, and derive the neural layer as the finite-dimensional input-output map of the reduced
solitonic dynamics. In this construction, the computational unit is a projected collective coordinate
of a localized field configuration rather than an elementary point variable; the activation function is
the solitonic response profile or scattering map; the weight matrix is the Hessian or overlap matrix
of an effective interaction energy among solitons; the bias is induced by external sources, vacuum
asymmetry, or boundary forcing; and depth is a discrete evolution parameter on the solitonic moduli
space. We develop the construction explicitly for the \(\phi^4\) kink, where the \(\tanh\) activation
and the logistic sigmoid arise from the kink profile, and then derive the multilayer feedforward
form from an operator-splitting approximation to collective-coordinate gradient flow. We emphasize
the novelty criterion: the neural architecture is obtained only after specifying the field action, the
solitonic ansatz, the moduli-space metric, the interaction functional, and the projection map. The
result is a controlled route from nonlinear field theory to neural-network structure, with robustness
tied to the energetic and topological stability of the solitonic sector.
\end{abstract}

\maketitle
\section{Introduction and statement of the construction}\label{sec:intro}

Artificial neural networks are normally defined algebraically as compositions of affine transformations and pointwise nonlinear maps.  A standard feedforward layer is written as
\begin{equation}
    h^{(\ell+1)}=\sigma\!\left(W^{(\ell)}h^{(\ell)}+b^{(\ell)}\right),
    \label{eq:standard_layer_intro}
\end{equation}
where $h^{(\ell)}$ is a finite-dimensional hidden state, $W^{(\ell)}$ is a matrix of trainable weights, $b^{(\ell)}$ is a bias, and $\sigma$ is an activation function; this representation is mathematically efficient and underlies much of modern machine learning theory, including approximation theorems, dynamical-systems interpretations, mean-field limits, and kernel limits~\cite{Cybenko1989,Hornik1991,Rosenblatt1958,Rumelhart1986,LeCun2015,Goodfellow2016,Chen2018,Jacot2018,MeiMontanariNguyen2018,BordelonPehlevan2022}.  From the standpoint of field theory, however, Eq.~\eqref{eq:standard_layer_intro} is a final reduced object: it contains localized units, nonlinear response, couplings, and discrete evolution, but it does not tell us whether such ingredients can arise from a more primitive continuum dynamics.

The question addressed here is therefore not whether neural networks can be compared with nonlinear field theories.  Such comparisons are abundant and useful: deep networks have been related to renormalization-group transformations, residual networks to differential equations, infinite-width networks to Gaussian processes and neural tangent kernels, and physics-informed neural networks to variational or differential constraints imposed by physical laws~\cite{MehtaSchwab2014,Chen2018,Raissi2019,Jacot2018,Lee2018,BordelonPehlevan2022}.  The question is sharper and more constructive: can one start from a nonlinear field theory that supports solitons and obtain Eq.~\eqref{eq:standard_layer_intro} as an effective finite-dimensional map of its nonperturbative sector?

\begin{figure}[t]
\centering
\begin{tikzpicture}[node distance=6.5mm, every node/.style={font=\small}, box/.style={draw, rounded corners, align=center, inner sep=4pt, minimum width=5.2cm}, arr/.style={-{Latex[length=2mm]}, thick}]
\node[box] (a) {nonlinear field action\\$S[\phi]$};
\node[box, below=of a] (b) {choice of solitonic sector\\$\phi(x,t)\simeq\Phi(x;q(t))$};
\node[box, below=of b] (c) {collective-coordinate reduction\\$S[\phi]\to S_{\rm eff}[q]$};
\node[box, below=of c] (d) {effective interaction network\\$G_{ab}(q)$ and $U_{\rm eff}(q)$};
\node[box, below=of d] (e) {projected neural layer\\$h^{(\ell+1)}=\sigma(W^{(\ell)}h^{(\ell)}+b^{(\ell)})$};
\draw[arr] (a) -- node[right, xshift=1mm] {nonperturbative sector} (b);
\draw[arr] (b) -- node[right, xshift=1mm] {moduli projection} (c);
\draw[arr] (c) -- node[right, xshift=1mm] {overlap integrals} (d);
\draw[arr] (d) -- node[right, xshift=1mm] {readout and splitting} (e);
\end{tikzpicture}
\caption{Constructive route from nonlinear field theory to a neural layer.  The essential point is the direction of derivation: the neural architecture is not assumed and then relabeled, but obtained after projecting the solitonic sector of the field theory onto collective coordinates and readout variables.}
\label{fig:construction_flow}
\end{figure}
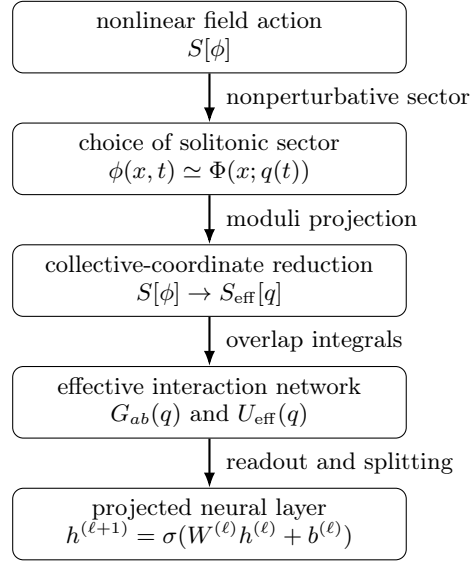

We answer this question affirmatively for a controlled class of models.  The proposed construction is summarized in Fig.~\ref{fig:construction_flow}.  The diagram is not intended as a decorative analogy; it fixes the order of operations used throughout the paper.  The field action is specified first, the solitonic sector is then selected, the collective-coordinate reduction is performed, the induced interaction network is computed, and only at the final stage does the neural layer appear as an effective map.
Figure~\ref{fig:construction_flow} is the central object of the paper.  Each arrow corresponds to a standard mathematical operation in nonlinear field theory rather than to a metaphor: one chooses a field action, selects a nonperturbative sector, expands the field on collective coordinates, integrates over space, identifies the induced metric and interaction potential on moduli space, and finally projects the reduced dynamics onto finite-dimensional input-output variables~\cite{Rajaraman1982,MantonSutcliffe2004,Vachaspati2006,Shifman2012}.

The physical reason this route is plausible is that solitons already possess the ingredients required by neural computation.  They are localized finite-energy objects; they are stabilized by topology, boundary conditions, nonlinearity, dispersion, or a balance of these mechanisms; they carry moduli such as position, width, phase, amplitude, and internal orientation; and they interact through overlap tails, external sources, nonlinear couplings, and lattice-mediated forces~\cite{DrazinJohnson1989,Rajaraman1982,MantonSutcliffe2004,DauxoisPeyrard2006,KartashovMalomedTorner2011,Malomed2016}.  The simplest $\phi^4$ kink has a profile proportional to $\tanh z$, which is one of the canonical neural activation functions, while its normalized form gives the logistic sigmoid; nevertheless, the identification of $\tanh$ with a kink is not the novelty of the present work.  The novelty is the derivation of the entire layer structure from a collective-coordinate reduction of interacting solitons.

The construction also clarifies what would count as a genuine contribution and what would not.  It would be merely a change of language to take a pretrained network and call its neurons ``solitons'' without specifying a field action, a solitonic ansatz, and a projection from field configurations to neural variables.  It becomes a field-theoretic construction only when the map
\begin{equation}
    \{S[\phi],\;\Phi(x;q),\;G_{ab}(q),\;U_{\rm eff}(q),\;\Pi\}
    \quad\Longrightarrow\quad
    \{\sigma,\;W,\;b,\;h^{(\ell)}\}
    \label{eq:novelty_map}
\end{equation}
can be written explicitly.  Here $\Pi$ denotes the projection from the continuum field or its moduli to the finite neural state.  In the minimal model developed below, $\sigma$ is inherited from the kink profile, $G_{ab}$ is the moduli-space metric, $W$ is an overlap/Hessian matrix of the solitonic interaction energy, $b$ is an induced source term, and the layer index is a discretized evolution parameter.

The article is organized to make this logic explicit.  Section~\ref{sec:literature} positions the construction relative to soliton physics and neural-network theory.  Section~\ref{sec:field_theory} defines the nonlinear field-theoretic setting and reviews finite-energy sectors.  Section~\ref{sec:solitonic_unit} derives the solitonic computational unit from the $\phi^4$ kink and explains the emergence of $\tanh$ and sigmoid activations.  Section~\ref{sec:collective} performs the collective-coordinate reduction.  Section~\ref{sec:weights} derives effective weights and biases from interaction energies.  Section~\ref{sec:layers} obtains a feedforward layer as a discrete solitonic evolution map.  Section~\ref{sec:learning} interprets learning as deformation and relaxation of the effective potential.  Section~\ref{sec:stability} discusses robustness and generalization in terms of Hessian stability and topological sectors.  Section~\ref{sec:minimal_model} gives a minimal worked construction.  Section~\ref{sec:applications} gives two graphical worked constructions.  Section~\ref{sec:phenomenology} discusses physical implications, limitations, and experimental directions.  Appendices give step-by-step derivations used in the main text.

\section{Relation to existing literature and novelty}\label{sec:literature}

\subsection{Solitons and nonlinear localized structures}

The literature on solitons is broad because localized nonlinear structures arise in many distinct physical regimes: relativistic scalar-field models, integrable wave equations, nonlinear optics, Bose--Einstein condensates, Josephson systems, discrete lattices, condensed-matter domain walls, and cosmological defect models~\cite{ZabuskyKruskal1965,FaddeevTakhtajan1987,DrazinJohnson1989,Rajaraman1982,MantonSutcliffe2004,Vachaspati2006,DauxoisPeyrard2006,KartashovMalomedTorner2011,Malomed2016}.  The common feature is not the detailed form of the equation, but the presence of stable or long-lived localized degrees of freedom whose collective variables evolve more slowly than their internal profile modes.  This separation between profile and moduli is the physical mechanism that makes a finite-dimensional reduction meaningful.

Solitons are among the most studied nonperturbative objects in nonlinear physics.  In relativistic field theory they include kinks, domain walls, vortices, monopoles, Skyrmions, instantons, and Q-balls; in nonlinear optics, condensed matter, Bose-Einstein condensates, and nonlinear lattices, related localized structures arise from balances among dispersion, nonlinearity, discreteness, and external potentials~\cite{Rajaraman1982,DrazinJohnson1989,MantonSutcliffe2004,Vachaspati2006,Shifman2012,DauxoisPeyrard2006,KartashovMalomedTorner2011}.  The theory of collective coordinates is also standard: one approximates a field configuration by a family of solitonic profiles parametrized by moduli, substitutes the ansatz into the action, and obtains a finite-dimensional effective dynamics~\cite{MantonSutcliffe2004,Vachaspati2006,Weinberg2012}.

The present work uses this established machinery but asks a different question.  Instead of treating collective coordinates as approximate particle coordinates for soliton scattering, quantization, or moduli-space dynamics, we treat them as computational degrees of freedom.  The induced moduli-space metric and interaction energy become the geometric and coupling data of a neural architecture.  In this sense the paper does not propose a new soliton solution; it proposes a new use of the solitonic sector as a constructive origin for neural layers.

The works of Malomed and collaborators on multidimensional solitons, nonlinear lattices, nonlocal nonlinear media, and stability mechanisms are relevant because they show how localized coherent states can persist and interact in structured nonlinear environments~\cite{Malomed2016,KartashovMalomedTorner2011,Malomed2022}.  The works of Correa, Dutra, Frederico, Malomed, Oliveira, and Sawado on oscillons and oscillating kinks in coupled scalar-field theories are relevant because they illustrate how controlled nonlinear potentials can generate families of localized long-lived or topological configurations with tunable shape parameters~\cite{CorreaDutra2016,CorreaDutraFredericoMalomed2019}.  These studies motivate the view that a neural unit need not be restricted to a static kink; it may also be a localized response or oscillatory solitonic mode with a stable collective description.

\subsection{Neural networks, continuum limits, and field-theoretic viewpoints}

The neural-network literature is equally broad, spanning classical perceptrons, universal approximation, backpropagation, convolutional and deep architectures, residual networks, infinite-width limits, and dynamical-system formulations~\cite{Rosenblatt1958,Cybenko1989,Hornik1991,Rumelhart1986,Goodfellow2016,LeCun2015,He2016,Chen2018,Jacot2018}.  These results explain why networks are expressive and trainable, but they do not by themselves give a microscopic physical origin for the layer map.  The present construction is therefore complementary to approximation theory and statistical learning theory: it asks what kind of nonlinear field dynamics can generate the algebraic form used by a feedforward network.

The machine-learning literature contains several continuum and field-theoretic perspectives, but they differ from the present construction.  The renormalization-group interpretation of deep learning relates hierarchical feature extraction to coarse-graining transformations and variational RG ideas~\cite{MehtaSchwab2014,Beny2013,LinTegmarkRolnick2017}.  Neural ordinary differential equations and PDE-inspired networks view depth as a time-like variable and residual blocks as discretizations of continuous flows~\cite{Chen2018,HaberRuthotto2017,RuthottoHaber2020}.  Infinite-width limits lead to Gaussian processes, neural tangent kernels, and mean-field dynamics~\cite{Neal1996,Lee2018,Jacot2018,MeiMontanariNguyen2018}.  Dynamical field theories for feature learning provide reduced descriptions in terms of order parameters and kernels that evolve during training~\cite{BordelonPehlevan2022}.

These approaches usually start from a neural network and derive a continuum, statistical, or field-theoretic description of that network.  The direction of the present work is reversed.  We start from a nonlinear field theory and derive a neural map from the solitonic sector.  This reversal matters.  In the standard continuum-limit program the field is an effective description of many neurons; here the neuron is an effective description of a localized field configuration.  The two viewpoints are compatible, but they are not identical.

\subsection{Precise novelty criterion}

To avoid overclaiming, we formulate the novelty criterion as a set of required ingredients.  A solitonic construction of a neural architecture must specify:
\begin{enumerate}[label=(\roman*)]
    \item a nonlinear field theory $S[\phi]$ with stable localized solutions;
    \item a solitonic ansatz $\Phi(x;q)$ with collective coordinates $q^a$;
    \item the reduced metric $G_{ab}(q)$ and effective potential $U_{\rm eff}(q)$ obtained by substitution into the action;
    \item a projection $\Pi$ from collective coordinates to neural states $h_i$;
    \item an approximation under which the reduced evolution becomes a layer map of the form Eq.~\eqref{eq:standard_layer_intro}.
\end{enumerate}
If these ingredients are absent, the discussion is only analogical.  If they are present, the neural architecture is a finite-dimensional effective theory of the solitonic sector.  This paper gives the construction explicitly for kink-like units and then indicates how the same logic extends to oscillons, domain walls, and nonlinear-lattice solitons.

\subsection{What is not new, and what is new}

Several individual observations used in this work are not new.  Kinks in the $\phi^4$ model have $\tanh$ profiles; collective coordinates reduce soliton dynamics to finite-dimensional mechanics; interaction energies generate effective forces; and feedforward neural networks use affine maps followed by nonlinear activations~\cite{Rajaraman1982,MantonSutcliffe2004,Vachaspati2006,Cybenko1989,Goodfellow2016}.  Treating any one of these statements as the whole paper would indeed amount to repackaging known material.

The new step is the composition of these facts into a single derivational chain with no neural layer assumed at the beginning.  In the construction below, the field theory is specified before the network; the soliton profile is solved before the activation is named; the moduli-space metric is calculated before the parameter geometry is invoked; and the effective interaction Hessian is expanded before the weights are introduced.  This ordering is essential.  A symbolic replacement such as ``neuron equals soliton'' would be empty; a map from $S[\phi]$ and $\Phi(x;q)$ to $\sigma$, $W$, $b$, and $h$ is substantive because it leads to explicit overlap integrals, reduced equations of motion, and quantitative checks against the original field dynamics.

A second point of novelty is that the construction distinguishes three levels often conflated in heuristic discussions.  The microscopic level is the continuum field action.  The mesoscopic level is the solitonic moduli space.  The computational level is the projected neural state.  The neural network lives only at the third level, and its validity depends on the separation of scales that allows the second level to close approximately under the dynamics.  This separation-of-scales requirement is standard in soliton physics, but it provides a concrete criterion for when a solitonic neural network is well-defined.

The phrase ``solitonic construction of artificial neural networks'' is therefore used in a restrictive sense.  It does not claim that every artificial network realized in software is literally a soliton system.  It claims that a well-defined class of neural architectures can be generated from nonlinear field theories by the same reduction methods used to describe the dynamics of kinks, oscillons, breathers, and localized modes.  The paper is consequently a construction principle rather than a phenomenological metaphor.

\section{Nonlinear field theory and finite-energy sectors}\label{sec:field_theory}

\subsection{Action, equations of motion, and boundary conditions}

We begin with a real scalar field $\phi$ on $(1+d)$-dimensional Minkowski spacetime,
\begin{equation}
    S[\phi]=\int \dd t\,\dd^d x\,\cL,
    \qquad
    \cL=\frac{1}{2}\partial_\mu\phi\,\partial^\mu\phi - V(\phi),
    \label{eq:scalar_action}
\end{equation}
where $V(\phi)$ is a nonlinear potential.  The Euler-Lagrange equation is
\begin{equation}
    \partial_\mu\partial^\mu\phi + V_\phi(\phi)=0,
    \qquad V_\phi\equiv \frac{\dd V}{\dd\phi}.
    \label{eq:eom}
\end{equation}
For static configurations in one spatial dimension,
\begin{equation}
    \frac{\dd^2\phi}{\dd x^2}=V_\phi(\phi),
    \label{eq:static_eom}
\end{equation}
with energy
\begin{equation}
    E[\phi]=\int_{-\infty}^{\infty}\dd x
    \left[\frac{1}{2}\left(\frac{\dd\phi}{\dd x}\right)^2+V(\phi)\right].
    \label{eq:energy}
\end{equation}
Finite energy requires that $\phi$ approach vacuum values of $V$ at spatial infinity,
\begin{equation}
    \phi(-\infty)=v_-,\qquad \phi(+\infty)=v_+,
    \qquad V(v_\pm)=0.
    \label{eq:finite_energy_bc}
\end{equation}
If the vacuum manifold has disconnected components, configurations with different endpoint data cannot be continuously deformed into one another while maintaining finite energy~\cite{Rajaraman1982,MantonSutcliffe2004,Vachaspati2006}.  In one dimension a topological charge can be written as
\begin{equation}
    Q=\frac{\phi(+\infty)-\phi(-\infty)}{2v}
    \label{eq:topological_charge}
\end{equation}
for a double-well model with vacua $\pm v$.

The boundary data are central to the proposed neural construction.  A solitonic unit is not merely a localized bump; it is a field configuration constrained by global endpoint conditions or by an energetic barrier.  This makes it a robust carrier of information.  In the reduced description, the information is not stored in the value of the field at a single point but in collective features of a localized configuration.  The distinction is important for the physical interpretation of a neural unit: a pointwise activation can fluctuate locally, whereas a topological or quasi-topological solitonic profile requires a coherent deformation of an extended field configuration.  This is the first reason why a solitonic construction can naturally encode robustness rather than merely nonlinearity~\cite{Rajaraman1982,MantonSutcliffe2004,Vachaspati2006,Weinberg2012}.

\subsection{Bogomolny decomposition and first-order profiles}

Suppose that the potential can be written in terms of a superpotential $W(\phi)$,
\begin{equation}
    V(\phi)=\frac{1}{2}W_\phi^2.
    \label{eq:superpotential}
\end{equation}
Then the energy is
\begin{align}
    E
    &=\frac{1}{2}\int\dd x\left(\phi_x^2+W_\phi^2\right) \\
    &=\frac{1}{2}\int\dd x\left(\phi_x\mp W_\phi\right)^2
    \pm \int\dd x\,\phi_x W_\phi \\
    &\geq \left|W(\phi(+\infty))-W(\phi(-\infty))\right|.
    \label{eq:bps_bound}
\end{align}
The bound is saturated by first-order equations
\begin{equation}
    \phi_x=\pm W_\phi.
    \label{eq:bps_equation}
\end{equation}
The first-order form is important because it already has the structure of a nonlinear response law.  The profile $\phi(x)$ is obtained by integrating a local nonlinear differential equation, and its shape is controlled by the potential.  In the neural construction, the activation function is precisely this response profile written in a dimensionless coordinate.

\subsection{From field configurations to computational units}
\label{subsec:field_configurations_to_computational_units}

A conventional artificial neuron is usually introduced as a scalar nonlinear response to an effective input,
\begin{equation}
h=\sigma(z),
\label{eq:standard_scalar_neuron}
\end{equation}
where \(z\) is an affine combination of previous computational states and \(\sigma\) is a prescribed activation function. In this standard formulation, the scalar variable \(h\) is taken as primitive: one postulates a finite-dimensional computational unit and then assigns to it a nonlinear response. The present construction reverses this logic. The scalar unit is not assumed at the outset. It is obtained by projecting a localized field configuration onto a finite-dimensional observable.

A solitonic unit is a localized nonlinear solution
\begin{equation}
\Phi(x;q),
\label{eq:solitonic_unit_profile}
\end{equation}
where \(q\) denotes one or more collective coordinates. Depending on the underlying field theory, these coordinates may describe the soliton position, phase, amplitude, width, internal orientation, separation from another soliton, or deformation along a zero mode. In this sense, the elementary object is not initially a neuron, but a coherent field configuration with a small number of physically meaningful degrees of freedom. The computational variable arises only after one specifies how this field configuration is read out.

To obtain a neuron from a soliton, one must choose a projection
\begin{equation}
h=\Pi[\Phi(x;q)],
\label{eq:projection_to_neuron}
\end{equation}
where \(h\) is the finite-dimensional computational state associated with the solitonic configuration. The map \(\Pi\) is a readout functional from the infinite-dimensional field configuration space to a finite-dimensional state space. Possible projections include the field value at a detector point,
\begin{equation}
h=\Phi(x_d;q),
\label{eq:point_detector_projection}
\end{equation}
an overlap with a localized test function,
\begin{equation}
h=
\int d^d x\,\chi(x)\Phi(x;q),
\label{eq:overlap_projection}
\end{equation}
the amplitude of a zero mode,
\begin{equation}
h=
\int d^d x\,
\eta_0(x;q)\,\delta\Phi(x),
\label{eq:zero_mode_projection}
\end{equation}
or a coarse-grained label associated with the asymptotic vacuum sector,
\begin{equation}
h=
\Pi_{\rm vac}[\Phi].
\label{eq:vacuum_label_projection}
\end{equation}
These alternatives are physically distinct. A point detector measures a local field value; an overlap functional measures how much the soliton resembles a chosen mode; a zero-mode projection extracts a collective deformation; and an asymptotic projection records the sector in which the configuration lies. Thus, different choices of \(\Pi\) generate different computational units.

The simplest choice used below is to identify the dimensionless input \(z\) with the translated kink coordinate and the output with the normalized kink profile. For a one-dimensional kink centered at \(X\), one writes
\begin{equation}
z=\mu(x-X),
\label{eq:dimensionless_kink_coordinate}
\end{equation}
where \(\mu^{-1}\) is the characteristic width of the wall. The normalized readout is then
\begin{equation}
h(z)=\frac{\Phi_K(x;X)}{v}.
\label{eq:normalized_kink_readout}
\end{equation}
For the \(\phi^4\) kink,
\begin{equation}
\Phi_K(x;X)
=
v\tanh[\mu(x-X)],
\label{eq:kink_profile_readout}
\end{equation}
so that
\begin{equation}
h(z)=\tanh z.
\label{eq:tanh_activation_from_projection}
\end{equation}
In this case the familiar hyperbolic tangent activation is not inserted by hand. It is the finite-dimensional readout of a topological field configuration. A sigmoid activation is obtained by shifting and normalizing the same profile,
\begin{equation}
h_{\rm sig}(z)
=
\frac{1}{2}\left[1+\tanh z\right].
\label{eq:sigmoid_activation_from_projection}
\end{equation}

This step is where the construction differs from a linguistic analogy. The neural scalar \(h\) is not simply renamed as a field value. It is defined as a functional of a localized field configuration. The activation function is likewise not merely compared with a soliton profile; it appears after choosing a physically meaningful projection of the soliton onto a reduced computational coordinate. In other words, the sequence is
\begin{equation}
\Phi(x;q)
\quad
\xrightarrow{\;\Pi\;}
\quad
h
\quad
\xrightarrow{\;\text{solitonic response}\;}
\quad
\sigma(z),
\label{eq:field_to_activation_sequence}
\end{equation}
rather than the reverse. The artificial neuron is therefore an effective degree of freedom extracted from the solitonic sector of the field theory.

The physical interpretation is straightforward. A soliton interpolates between distinct asymptotic regimes of the field. For the kink, these regimes are the two vacua \(\phi=-v\) and \(\phi=+v\). When the detector coordinate \(z\) is far to one side of the soliton center, the projected value saturates near one vacuum. When \(z\) is far to the other side, it saturates near the other vacuum. The nonlinear transition between these two limits is precisely the activation curve. Thus, saturation in the neural unit corresponds to the approach of the field configuration to a vacuum sector, while the nonlinear transition region corresponds to the core of the soliton.

This observation also clarifies why solitonic units are natural computational objects. A generic field fluctuation disperses, decays, or mixes with other modes. A soliton, by contrast, is a coherent nonlinear excitation whose profile is protected by the structure of the field equations and, in topological cases, by boundary conditions. Consequently, the projected variable \(h=\Pi[\Phi]\) can remain stable under small microscopic perturbations of the field. This provides a physical mechanism for robust finite-dimensional computation: the computational state is not attached to an arbitrary microscopic degree of freedom, but to a collective excitation with structural stability.

Different choices of projection lead to different neural architectures, just as different choices of collective coordinates lead to different effective moduli-space dynamics. A detector projection naturally produces a local activation unit. An overlap projection produces a mode-selective unit, closer to a feature detector. A projection onto relative coordinates between two solitons produces a unit sensitive to interaction and separation. A projection onto internal orientations produces a vector-valued or multi-channel unit. Therefore the construction does not identify a soliton with a neuron in a unique or universal way. Rather, it provides a systematic field-theoretic mechanism for generating computational units from localized nonlinear configurations.

The role of the projection can be summarized as follows:
\begin{equation}
\boxed{
\text{field configuration}
\;\longrightarrow\;
\text{solitonic collective coordinate}
\;\longrightarrow\;
\text{projected computational state}
\;\longrightarrow\;
\text{neural response}
}
\label{eq:field_to_neuron_summary}
\end{equation}
This is the first essential ingredient of the solitonic construction. The second ingredient, developed in the following subsection, is the emergence of effective weights from interactions among several projected solitonic units.

\section{The solitonic neuron: kink profiles as activations}\label{sec:solitonic_unit}

\subsection{Derivation of the $\phi^4$ kink}

Consider the $\phi^4$ double-well theory
\begin{equation}
    V(\phi)=\frac{\lambda}{4}(\phi^2-v^2)^2,
    \qquad \lambda>0,
    \qquad v>0.
    \label{eq:phi4_potential}
\end{equation}
The vacua are $\phi=\pm v$.  The static equation is
\begin{equation}
    \phi_{xx}=\lambda\phi(\phi^2-v^2).
    \label{eq:phi4_static}
\end{equation}
A compatible superpotential derivative is
\begin{equation}
    W_\phi=\sqrt{\frac{\lambda}{2}}(v^2-\phi^2),
    \label{eq:phi4_Wphi}
\end{equation}
so that Eq.~\eqref{eq:bps_equation} gives
\begin{equation}
    \frac{\dd\phi}{\dd x}=\sqrt{\frac{\lambda}{2}}(v^2-\phi^2).
    \label{eq:phi4_first_order}
\end{equation}
Separating variables,
\begin{equation}
    \int\frac{\dd\phi}{v^2-\phi^2}
    =\sqrt{\frac{\lambda}{2}}\int\dd x.
    \label{eq:phi4_integral}
\end{equation}
Using
\begin{equation}
    \int\frac{\dd\phi}{v^2-\phi^2}
    =\frac{1}{2v}\ln\left(\frac{v+\phi}{v-\phi}\right),
    \label{eq:phi4_integral_eval}
\end{equation}
we obtain
\begin{equation}
    \frac{1}{2v}\ln\left(\frac{v+\phi}{v-\phi}\right)
    =\sqrt{\frac{\lambda}{2}}(x-X),
    \label{eq:phi4_log}
\end{equation}
where $X$ is an integration constant.  Solving for $\phi$ yields
\begin{equation}
    \phi_K(x;X)=v\tanh\left[\frac{m}{2}(x-X)\right],
    \qquad m=\sqrt{2\lambda}v.
    \label{eq:phi4_kink}
\end{equation}
The kink is a topological soliton interpolating between $-v$ and $+v$, and $X$ is its translational collective coordinate~\cite{Rajaraman1982,MantonSutcliffe2004,Vachaspati2006}.

\subsection{Activation as a dimensionless solitonic response}

Define the dimensionless coordinate
\begin{equation}
    z=\frac{m}{2}(x-X).
    \label{eq:z_def}
\end{equation}
The normalized kink profile is then
\begin{equation}
    a(z)=\frac{\phi_K}{v}=\tanh z.
    \label{eq:tanh_from_kink}
\end{equation}
Thus the $\tanh$ activation is not introduced as an arbitrary nonlinearity; it is the normalized response of a topological soliton.  The logistic sigmoid follows by affine normalization:
\begin{equation}
    s(z)=\frac{1}{2}\left[1+\tanh z\right]
    =\frac{1}{1+e^{-2z}}.
    \label{eq:sigmoid_from_kink}
\end{equation}
The factor of $2$ in the exponential is conventional and can be absorbed into the scale defining $z$.  The key point is that both activations arise from the same kink profile.

\begin{figure}[t]
\centering
\includegraphics[width=0.72\textwidth]{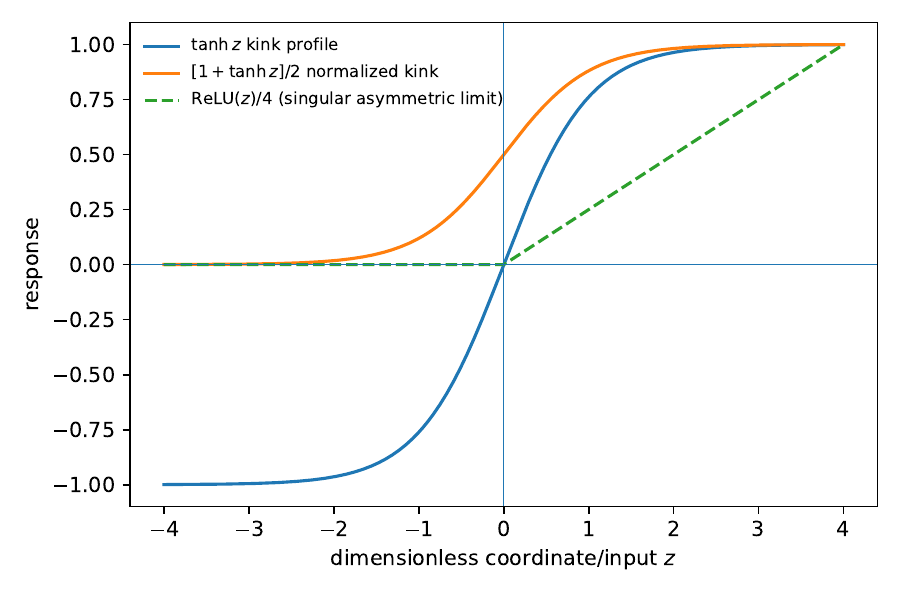}
\caption{Solitonic activation profiles.  The $\phi^4$ kink gives $\tanh z$ directly, while the normalized kink gives the logistic-type response $[1+\tanh z]/2$.  The ReLU curve is shown only as a singular asymmetric limiting response, not as a smooth topological kink.}
\label{fig:activation_profiles}
\end{figure}

This result alone is not enough to claim a solitonic origin of neural networks.  It shows only that a familiar activation can be realized as a field profile.  The stronger claim requires the preactivation $z$ to arise from interactions among solitonic modes.  In a neural layer, $z_i=\sum_j W_{ij}h_j+b_i$.  In the solitonic construction, this combination must be derived from an effective force, overlap, or source acting on the $i$th localized mode.  This is developed in Secs.~\ref{sec:collective} and~\ref{sec:weights}.

\subsection{General solitonic activations}

The $\phi^4$ kink gives $\tanh$, but the construction is not limited to this model.  A sine-Gordon kink gives
\begin{equation}
    \phi_{\rm SG}(x)=4\arctan e^{m(x-X)},
    \label{eq:sine_gordon_kink}
\end{equation}
whose normalized response is another smooth monotone transition between vacua~\cite{DrazinJohnson1989,Rajaraman1982}.  Deformed scalar-field models can generate compactons, asymmetric kinks, flat-top kinks, and multi-kink profiles~\cite{Bazeia2002,CorreaDutraFredericoMalomed2019}.  Oscillons provide time-dependent localized responses whose envelope or phase may play the role of a computational output~\cite{Gleiser1994,CorreaDutra2016}.  Nonlocal nonlinear media can produce response functions whose output depends on an integral kernel rather than a strictly local preactivation~\cite{Malomed2022}.

The general recipe is therefore
\begin{equation}
    \sigma(z)=\Pi\left[\Phi_{\rm sol}(x;q(z))\right],
    \label{eq:general_solitonic_activation}
\end{equation}
where $q(z)$ is the collective coordinate driven by the effective input.  This equation defines a class of solitonic activations.  Standard $\tanh$ and sigmoid are the simplest examples, but the field-theoretic origin can generate nonlinearities with controlled saturation, asymmetry, compact support, oscillatory tails, or nonlocal response.

\section{Collective-coordinate reduction}\label{sec:collective}

\subsection{Moduli-space ansatz}

Let $\Phi(x;q)$ be a family of solitonic configurations parametrized by collective coordinates $q^a(t)$, $a=1,\ldots,n$.  We approximate the full field by
\begin{equation}
    \phi(x,t)=\Phi(x;q(t))+\eta(x,t),
    \label{eq:field_decomposition}
\end{equation}
where $\eta$ denotes fluctuations orthogonal to the chosen collective modes.  At leading order we set $\eta=0$ and substitute the ansatz into the action.  The time derivative is
\begin{equation}
    \dot\phi(x,t)=\frac{\partial \Phi}{\partial q^a}\dot q^a,
    \label{eq:time_derivative_moduli}
\end{equation}
where repeated indices are summed.  The kinetic term becomes
\begin{equation}
    T_{\rm eff}=\frac{1}{2}G_{ab}(q)\dot q^a\dot q^b,
    \qquad
    G_{ab}(q)=\int\dd^d x\,
    \frac{\partial\Phi}{\partial q^a}
    \frac{\partial\Phi}{\partial q^b}.
    \label{eq:moduli_metric}
\end{equation}
The potential term gives
\begin{equation}
    U_{\rm eff}(q)=\int\dd^d x\left[
    \frac{1}{2}(\nabla\Phi)^2+V(\Phi)
    \right].
    \label{eq:effective_potential}
\end{equation}
Thus the reduced action is
\begin{equation}
    S_{\rm eff}[q]=\int\dd t\left[
    \frac{1}{2}G_{ab}(q)\dot q^a\dot q^b-U_{\rm eff}(q)
    \right].
    \label{eq:effective_action}
\end{equation}
This is a standard collective-coordinate reduction, but here $q^a$ will be projected to neural states.

\subsection{Effective equations and dissipative relaxation}

The Euler-Lagrange equations of Eq.~\eqref{eq:effective_action} are
\begin{equation}
    G_{ab}\ddot q^b+\Gamma_{abc}\dot q^b\dot q^c+\partial_a U_{\rm eff}=0,
    \label{eq:moduli_eom}
\end{equation}
where
\begin{equation}
    \Gamma_{abc}=\frac{1}{2}\left(\partial_bG_{ac}+\partial_cG_{ab}-\partial_aG_{bc}\right).
    \label{eq:christoffel_lower}
\end{equation}
For learning and computation it is often useful to consider an overdamped or relaxational limit,
\begin{equation}
    \gamma_{ab}\dot q^b=-\partial_a U_{\rm eff}+J_a(t),
    \label{eq:overdamped}
\end{equation}
where $\gamma_{ab}$ is a positive friction tensor and $J_a$ is an external source.  If $\gamma_{ab}$ is identified with or proportional to the moduli-space metric, the flow becomes a natural gradient flow,
\begin{equation}
    \dot q^a=-G^{ab}\partial_b U_{\rm eff}+G^{ab}J_b.
    \label{eq:natural_gradient_flow}
\end{equation}
This equation is the continuum ancestor of the neural update.  The field theory supplies both the geometry $G_{ab}$ and the effective energy $U_{\rm eff}$.

\subsection{Projection to neural variables}

Let $h_i$ be the neural state associated with the $i$th solitonic unit.  We define
\begin{equation}
    h_i=\Pi_i[\Phi(x;q)]=\int\dd^d x\,\rho_i(x)\,\Phi(x;q),
    \label{eq:projection_rho}
\end{equation}
where $\rho_i(x)$ is a localized readout function.  Other projections are possible, but Eq.~\eqref{eq:projection_rho} is sufficient for the derivation.  If the $i$th soliton is centered near a detector point and has a kink-like response, then
\begin{equation}
    h_i\simeq \sigma(z_i),
    \label{eq:hi_sigma_zi}
\end{equation}
where $z_i$ is the effective coordinate controlling the displacement, source, or local field acting on the soliton.  The remaining task is to show that $z_i$ is an affine function of other neural states at leading order.

\subsection{Separation of scales and closure of the reduced description}

The collective-coordinate construction requires a separation between slow solitonic coordinates and fast radiation modes.  Let $\eta$ denote fluctuations orthogonal to the collective-coordinate tangent vectors.  The reduced neural description is reliable when the radiation energy carried by $\eta$ remains perturbative over the time interval used for the layer map,
\begin{equation}
    \frac{\|\eta\|_{\cH}^2}{E[\Phi(x;q)]}\ll 1,
    \label{eq:eta_small_condition}
\end{equation}
where $\|\eta\|_{\cH}^2=\int\eta\cH\eta\,\dd x$ is defined by the fluctuation operator.  If this condition fails, the solitonic object radiates strongly or changes its identity, and the finite-dimensional neural variable is no longer closed.  This is not a weakness of the construction; it is a quantitative regime condition.

The same condition has a computational interpretation.  A neural layer assumes that the state passed to the next layer is finite-dimensional.  In the solitonic picture, this means that the continuum field has been compressed into a small number of robust collective coordinates.  The validity of the compression is controlled by the spectral gap between zero or shape modes and the radiation continuum.  A gapped stable soliton therefore gives a more reliable neural unit than an unstable localized bump without a protected or metastable sector~\cite{Rajaraman1982,MantonSutcliffe2004,Vachaspati2006}.

\section{Weights and biases from solitonic interactions}\label{sec:weights}

\subsection{Interaction energy and overlap matrix}

Consider $N$ localized solitonic modes.  A useful ansatz is
\begin{equation}
    \Phi(x;q)=\sum_{i=1}^{N} A_i\,\Phi_i(x-X_i;\alpha_i)-\Phi_{\rm vac}^{\rm sub},
    \label{eq:multi_soliton_ansatz}
\end{equation}
where $X_i$ are centers, $A_i$ are amplitudes or orientations, $\alpha_i$ are shape parameters, and the vacuum subtraction avoids double counting the asymptotic background.  Substituting into the energy functional gives
\begin{equation}
    U_{\rm eff}(q)=\sum_i U_i(q_i)+\sum_{i<j}U_{ij}(q_i,q_j)+\sum_{i<j<k}U_{ijk}+\cdots.
    \label{eq:interaction_expansion}
\end{equation}
At large separation, the two-body terms are controlled by tail overlaps and typically decay exponentially for massive fields,
\begin{equation}
    U_{ij}\sim C_{ij}e^{-m|X_i-X_j|}
    \label{eq:tail_overlap}
\end{equation}
up to model-dependent signs and polynomial prefactors~\cite{MantonSutcliffe2004,DauxoisPeyrard2006}.  In nonlinear lattices or externally structured media, the couplings may also be shaped by the lattice profile or by a nonlocal response kernel~\cite{KartashovMalomedTorner2011,Malomed2022}.

Near a reference configuration $q_\star$, expand the effective potential:
\begin{equation}
    U_{\rm eff}(q)=U_\star+F_a\delta q^a+\frac{1}{2}K_{ab}\delta q^a\delta q^b+\frac{1}{3!}K_{abc}\delta q^a\delta q^b\delta q^c+\cdots,
    \label{eq:potential_expansion}
\end{equation}
where
\begin{equation}
    K_{ab}=\left.\frac{\partial^2U_{\rm eff}}{\partial q^a\partial q^b}\right|_{q_\star}.
    \label{eq:hessian_K}
\end{equation}
The off-diagonal components of $K_{ab}$ measure how the displacement of one solitonic mode changes the force on another.  These are precisely the ingredients that become effective weights after projection.

\begin{figure}[t]
\centering
\includegraphics[width=0.70\textwidth]{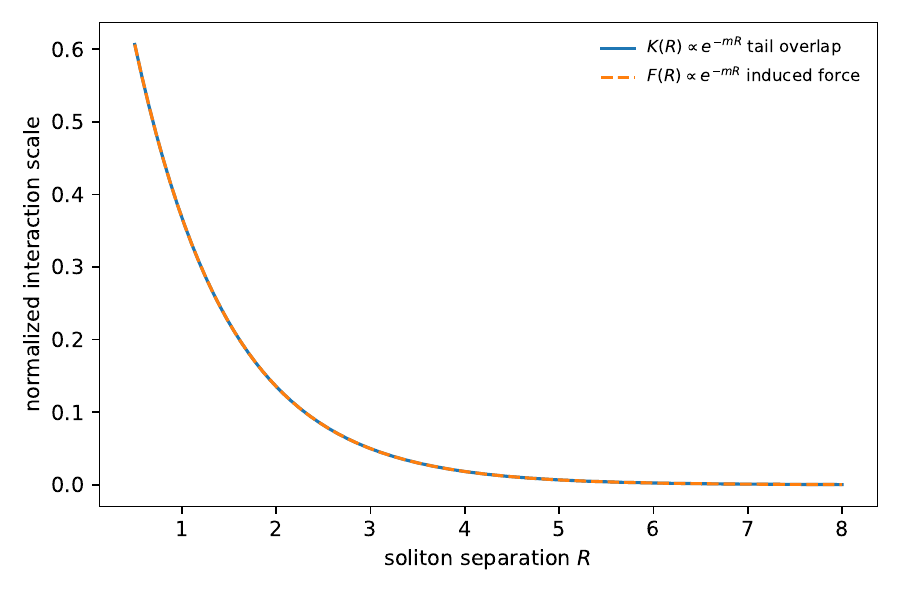}
\caption{Schematic solitonic origin of effective weights.  For massive fields, the tails of well-separated kinks typically generate exponentially suppressed overlaps.  After projection, these overlap scales contribute to the effective coupling matrix and therefore to the neural weights.}
\label{fig:overlap_weight}
\end{figure}

\subsection{Derivation of the affine preactivation}

Suppose the $i$th solitonic output is controlled by a coordinate $z_i$ satisfying a quasi-static balance equation
\begin{equation}
    \mu_i z_i + \sum_j K_{ij}h_j = J_i,
    \label{eq:balance_z}
\end{equation}
where $\mu_i>0$ is a local stiffness and $J_i$ is an external source.  Solving for $z_i$ gives
\begin{equation}
    z_i=\sum_j W_{ij}h_j+b_i,
    \qquad
    W_{ij}=-\frac{K_{ij}}{\mu_i},
    \qquad
    b_i=\frac{J_i}{\mu_i}.
    \label{eq:weights_biases}
\end{equation}
If the soliton response is $h_i=\sigma_i(z_i)$, then
\begin{equation}
    h_i=\sigma_i\left(\sum_jW_{ij}h_j+b_i\right).
    \label{eq:solitonic_neuron_equation}
\end{equation}
This is the neural unit equation derived from an effective solitonic balance condition.  The weight matrix is not postulated; it is a rescaled interaction Hessian or overlap matrix.  The bias is not postulated; it is a rescaled source, boundary tilt, or vacuum-asymmetry term.

The derivation also clarifies the regime of validity.  Equation~\eqref{eq:weights_biases} is leading order in a local expansion around a reference configuration.  Higher-order terms in Eq.~\eqref{eq:potential_expansion} generate higher-order neural interactions,
\begin{equation}
    z_i=\sum_jW_{ij}h_j+\sum_{jk}W_{ijk}^{(2)}h_jh_k+\cdots+b_i,
    \label{eq:higher_order_neuron}
\end{equation}
which may be interpreted as nonlinear synaptic couplings or higher-body solitonic interactions.  Ordinary feedforward networks correspond to the leading bilinear truncation.

\subsection{Sign, locality, and sparsity}

The sign of $W_{ij}$ depends on whether the relevant soliton-soliton force is effectively attractive or repulsive in the chosen projection.  The magnitude depends on overlap, separation, medium profile, and internal orientation.  Since massive-field soliton tails often decay exponentially, locality in physical space naturally induces sparsity in the effective network.  Conversely, nonlocal media or auxiliary mediator fields can generate long-range dense couplings~\cite{KartashovMalomedTorner2011,Malomed2022}.  This gives a physical interpretation to architectural priors: convolutional locality corresponds to translation-structured couplings, sparse networks correspond to short-range solitonic overlap, and dense layers correspond to globally mediated interactions.

\section{Feedforward layers as discrete solitonic evolution}\label{sec:layers}

\subsection{Continuous reduced flow}

Let $h(\tau)$ denote the projected neural state of a solitonic configuration at effective evolution parameter $\tau$.  The reduced overdamped dynamics can be written abstractly as
\begin{equation}
    \frac{\dd h}{\dd\tau}=F(h;\Theta),
    \label{eq:reduced_flow_h}
\end{equation}
where $\Theta$ denotes the field-theoretic parameters controlling the effective metric, potential, couplings, and sources.  In a residual form, a time step gives
\begin{equation}
    h^{(\ell+1)}=h^{(\ell)}+\Delta\tau\,F(h^{(\ell)};\Theta_\ell)+\cO(\Delta\tau^2).
    \label{eq:residual_step}
\end{equation}
This is closely related to the dynamical-systems interpretation of residual networks and neural ordinary differential equations~\cite{Chen2018,HaberRuthotto2017,RuthottoHaber2020}, but with a different origin: $F$ is not an arbitrary neural vector field; it is induced by solitonic collective-coordinate dynamics.

\subsection{Operator splitting and the standard layer}

To obtain the standard non-residual layer, separate the evolution into an interaction step and a response step.  First compute the effective input generated by interactions:
\begin{equation}
    z^{(\ell)}=W^{(\ell)}h^{(\ell)}+b^{(\ell)}.
    \label{eq:interaction_step}
\end{equation}
Then apply the local solitonic response profile:
\begin{equation}
    h^{(\ell+1)}=\sigma\left(z^{(\ell)}\right).
    \label{eq:response_step}
\end{equation}
Composing Eqs.~\eqref{eq:interaction_step} and~\eqref{eq:response_step} gives
\begin{equation}
    h^{(\ell+1)}=\sigma\left(W^{(\ell)}h^{(\ell)}+b^{(\ell)}\right),
    \label{eq:derived_layer}
\end{equation}
which is Eq.~\eqref{eq:standard_layer_intro}.  In this derivation $W^{(\ell)}$ and $b^{(\ell)}$ may change with $\ell$ because the effective medium, source profile, or interaction geometry changes along the evolution parameter.

The layer index therefore has a physical interpretation:
\begin{equation}
    \ell\quad\leftrightarrow\quad \text{discrete step in solitonic moduli-space evolution}.
    \label{eq:depth_interpretation}
\end{equation}
Depth is not simply a stack of arbitrary algebraic maps; it is a finite-time sequence of interaction-response operations in the reduced nonperturbative sector.

\subsection{What is new in the layer derivation}

The standard layer Eq.~\eqref{eq:derived_layer} is familiar.  The new content lies in the prehistory of its symbols.  In the solitonic construction,
\begin{align}
    \sigma &\leftarrow \text{profile or scattering response of a soliton}, \\
    W_{ij} &\leftarrow \text{overlap/Hessian coefficient of } U_{\rm eff}, \\
    b_i &\leftarrow \text{external source, boundary tilt, or vacuum asymmetry}, \\
    h_i &\leftarrow \text{projection of a localized field configuration}, \\
    \ell &\leftarrow \text{discrete evolution parameter on moduli space}.
    \label{eq:symbol_origins}
\end{align}
This dictionary is not an analogy if each arrow is computed from a chosen $S[\phi]$ and $\Phi(x;q)$.  The rest of the paper is devoted to showing how such computations are organized and what they imply.

\subsection{Layer locality, directionality, and effective causality}

A feedforward network is directed, whereas many solitonic interactions are reciprocal at the level of the underlying energy.  Directionality can enter the effective neural map in three standard ways.  First, one may use a time-ordered protocol in which one group of solitons is externally driven and another group is read out after relaxation.  Second, dissipative or open-system terms can break microscopic time-reversal symmetry in the reduced dynamics.  Third, a spatially structured medium can impose an ordering of interactions, as in wave propagation through layers of a nonlinear optical system.  In each case the underlying field theory supplies reciprocal or causal dynamics, while the computational protocol defines the directed layer map.

This point is important for the PRD-style framing of the work.  We are not claiming that a static Hamiltonian with symmetric Hessian automatically equals an arbitrary directed neural network.  Rather, the directed feedforward map arises after specifying a preparation, interaction, relaxation, and readout procedure.  The matrix $W^{(\ell)}$ is therefore an effective transfer matrix for a layer protocol, not necessarily a fundamental microscopic coupling matrix.

\section{Learning as deformation of the solitonic effective potential}\label{sec:learning}

\subsection{Data term as an external deformation}

Let $\{(x_n,y_n)\}_{n=1}^M$ be a supervised dataset.  A neural model computes $f_\theta(x_n)$ and minimizes a loss
\begin{equation}
    \cL_{\rm data}(\theta)=\sum_{n=1}^{M}\ell(f_\theta(x_n),y_n).
    \label{eq:data_loss}
\end{equation}
In the solitonic formulation, the parameters $\theta$ are effective coordinates or couplings inherited from the field theory.  The data term is interpreted as an external deformation of the reduced energy:
\begin{equation}
    U_{\rm total}(q)=U_{\rm eff}(q)+U_{\rm data}(q;\cD),
    \label{eq:total_energy_data}
\end{equation}
where $\cD$ denotes the dataset.  Training is then a relaxation process on moduli space,
\begin{equation}
    \frac{\dd q^a}{\dd s}=-G^{ab}(q)\frac{\partial U_{\rm total}}{\partial q^b},
    \label{eq:training_natural_gradient}
\end{equation}
with $s$ a training time.  This is a natural-gradient-like flow because the metric is induced by the field configuration itself~\cite{Amari1998,MantonSutcliffe2004}.

\begin{figure}[t]
\centering
\includegraphics[width=0.70\textwidth]{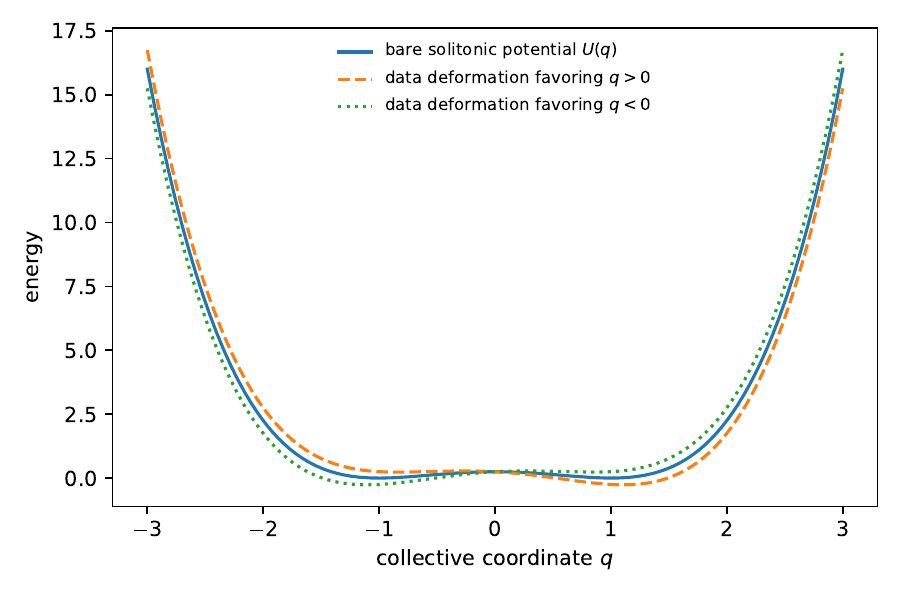}
\caption{Learning as deformation of an effective solitonic potential.  The bare collective-coordinate potential contains the intrinsic solitonic sector.  A data-dependent term tilts or reshapes this energy, selecting the collective configuration whose readout best matches the supervised target.}
\label{fig:energy_deformation}
\end{figure}

This interpretation differs from physics-informed neural networks.  In physics-informed learning, one usually trains a neural network while penalizing violations of known differential equations~\cite{Raissi2019}.  Here the differential equation precedes the neural network.  The network is the reduced solitonic dynamics, and the data deform the effective potential that selects which solitonic configuration is realized.

\subsection{Training weights as changing the medium}

If $W_{ij}$ is an interaction coefficient, training the weight matrix corresponds physically to changing the effective medium, the separation between solitonic modes, the strength of mediator fields, or the local parameters of the nonlinear potential.  A trainable weight is therefore not fundamental in the continuum theory; it is a controllable parameter of the reduced interaction functional.  For example, if
\begin{equation}
    U_{ij}=C_{ij}(\theta)e^{-m|X_i-X_j|}h_ih_j,
    \label{eq:trainable_interaction}
\end{equation}
then changing $C_{ij}$ or $X_i-X_j$ changes the induced weight.  This observation suggests a physical implementation principle: learning can be realized by adjusting background fields, lattice depths, refractive-index profiles, coupling channels, or boundary sources rather than by directly modifying abstract matrix entries.

\subsection{Backpropagation and adjoint dynamics}

The gradient of a multilayer map can be computed by the chain rule.  In continuous-depth systems, the corresponding object is an adjoint equation~\cite{Chen2018}.  In the solitonic construction, the same logic applies to the reduced moduli-space flow.  If
\begin{equation}
    \dot q^a=F^a(q;\Theta),
    \label{eq:q_flow}
\end{equation}
then the adjoint $p_a$ associated with a terminal loss satisfies
\begin{equation}
    \dot p_a=-p_b\frac{\partial F^b}{\partial q^a},
    \label{eq:adjoint_moduli}
\end{equation}
with additional terms for derivatives with respect to trainable medium parameters.  Thus backpropagation is not added by hand; it is the adjoint sensitivity equation of the reduced solitonic dynamics.

\section{Stability, robustness, and generalization}\label{sec:stability}

\subsection{Fluctuation operator around a soliton}

Let $\phi_{\rm sol}$ be a static solution.  Write
\begin{equation}
    \phi(x,t)=\phi_{\rm sol}(x)+\eta(x,t).
    \label{eq:fluctuation_ansatz}
\end{equation}
Expanding the energy to quadratic order gives
\begin{equation}
    E[\phi]=E[\phi_{\rm sol}]+\frac{1}{2}\int\dd x\,\eta\,\cH\,\eta+\cO(\eta^3),
    \label{eq:quadratic_energy}
\end{equation}
with fluctuation operator
\begin{equation}
    \cH=-\frac{\dd^2}{\dd x^2}+V_{\phi\phi}(\phi_{\rm sol}).
    \label{eq:fluctuation_operator}
\end{equation}
For the $\phi^4$ kink,
\begin{equation}
    V_{\phi\phi}(\phi_K)=\lambda(3\phi_K^2-v^2)
    =\frac{m^2}{2}\left(3\tanh^2\frac{m(x-X)}{2}-1\right),
    \label{eq:phi4_fluctuation_potential}
\end{equation}
which yields a Schr\"odinger-type stability problem with a translational zero mode and nonnegative spectrum for the stable kink sector~\cite{Rajaraman1982,Vachaspati2006}.

\subsection{Robustness as stability of a sector}

A learned solitonic network is robust when small perturbations of the input or of the medium do not move the configuration out of its stable sector.  In the reduced model this means that the Hessian of $U_{\rm total}$ in directions orthogonal to zero modes is positive,
\begin{equation}
    \delta^2U_{\rm total}=\frac{1}{2}H_{ab}\delta q^a\delta q^b>0
    \label{eq:positive_hessian_moduli}
\end{equation}
for allowed perturbations.  Topological stability is stronger: the configuration cannot be continuously unwound without changing boundary conditions or crossing an energetic barrier.  Neural generalization is not proven by this statement, but the statement provides a physical mechanism for robustness: classification or regression outputs associated with stable solitonic sectors are less sensitive to small deformations than outputs associated with unstable saddles.

This interpretation is deliberately modest.  It does not replace statistical learning theory, VC-dimension arguments, margin theory, PAC-Bayes analysis, or NTK/mean-field approaches~\cite{Goodfellow2016,Jacot2018,MeiMontanariNguyen2018}.  It adds a complementary field-theoretic criterion: a solitonic architecture is expected to be robust when the learned map corresponds to a stable basin or topological sector of the underlying nonlinear field theory.

\section{Minimal worked construction}\label{sec:minimal_model}

\subsection{Field theory and solitonic readout}

Take the one-dimensional $\phi^4$ theory of Eq.~\eqref{eq:phi4_potential}.  For a single unit, define the normalized readout
\begin{equation}
    h=\frac{1}{v}\phi_K(x_d;X)=\tanh\left[\frac{m}{2}(x_d-X)\right],
    \label{eq:single_readout}
\end{equation}
where $x_d$ is a detector coordinate.  Let
\begin{equation}
    z=\frac{m}{2}(x_d-X).
    \label{eq:single_z}
\end{equation}
Then $h=\tanh z$.  A source or interaction that shifts $X$ therefore changes the preactivation $z$.

\subsection{Two-unit interaction and induced weight}

Now consider two localized units with readouts $h_1$ and $h_2$.  Suppose their reduced interaction energy near a reference configuration is
\begin{equation}
    U_{\rm eff}=\frac{1}{2}\mu_1z_1^2+\frac{1}{2}\mu_2z_2^2+K_{12}h_1h_2-J_1z_1-J_2z_2.
    \label{eq:two_unit_energy}
\end{equation}
Quasi-static minimization with respect to $z_2$ gives
\begin{equation}
    \frac{\partial U_{\rm eff}}{\partial z_2}=\mu_2z_2+K_{12}h_1\frac{\partial h_2}{\partial z_2}-J_2=0.
    \label{eq:two_unit_balance_exact}
\end{equation}
In the weak-response or local-linear approximation $\partial h_2/\partial z_2\simeq c_2$, this becomes
\begin{equation}
    z_2=W_{21}h_1+b_2,
    \qquad
    W_{21}=-\frac{c_2K_{12}}{\mu_2},
    \qquad
    b_2=\frac{J_2}{\mu_2}.
    \label{eq:two_unit_weight}
\end{equation}
The output is therefore
\begin{equation}
    h_2=\tanh(W_{21}h_1+b_2).
    \label{eq:two_unit_neuron}
\end{equation}
This is the one-input neural unit derived from solitonic interaction.

\subsection{Network of $N$ units}

For $N$ units, choose an effective energy
\begin{equation}
    U_{\rm eff}=\sum_i\frac{1}{2}\mu_i z_i^2
    +\sum_{i<j}K_{ij}h_ih_j
    -\sum_iJ_iz_i.
    \label{eq:N_unit_energy}
\end{equation}
The same quasi-static balance gives
\begin{equation}
    z_i=\sum_jW_{ij}h_j+b_i,
    \qquad
    W_{ij}=-\frac{c_iK_{ij}}{\mu_i},
    \qquad
    b_i=\frac{J_i}{\mu_i}.
    \label{eq:N_unit_weight}
\end{equation}
Applying the solitonic response yields
\begin{equation}
    h_i=\tanh\left(\sum_jW_{ij}h_j+b_i\right).
    \label{eq:N_unit_neuron}
\end{equation}
A feedforward architecture is obtained by imposing a directed or time-ordered interaction pattern between layers,
\begin{equation}
    h_i^{(\ell+1)}=\tanh\left(\sum_jW_{ij}^{(\ell)}h_j^{(\ell)}+b_i^{(\ell)}\right).
    \label{eq:minimal_feedforward}
\end{equation}
This is a conventional neural layer in form, but every symbol has been inherited from the solitonic field theory.

\section{Two worked constructions and graphical interpretation}\label{sec:applications}

The previous sections establish the general derivation.  We now give two concrete constructions whose purpose is not to compete with applied machine-learning benchmarks, but to show how the field-theoretic ingredients translate into neural operations.  Both examples follow the same prescription: choose the nonlinear field model, identify the solitonic profile, define the readout, compute or approximate the effective interaction, and then read off the neural map.  This is the operational meaning of the title of the paper: a neural network is constructed from the solitonic sector rather than postulated independently.

\subsection{Example I: a one-kink binary response as a solitonic neuron}

Consider the $\phi^4$ model of Eq.~\eqref{eq:phi4_potential}.  Let the computational input be a dimensionless displacement variable
\begin{equation}
    z=\alpha x_{\rm in}+\beta,
\end{equation}
where $x_{\rm in}$ is an external scalar signal and $\alpha,\beta$ are source-dependent calibration constants.  The kink readout is defined by
\begin{equation}
    h(z)=\frac{1}{2}\left[1+\frac{\phi_K(z)}{v}\right]
    =\frac{1}{2}\left[1+\tanh z\right].
    \label{eq:one_kink_binary_readout}
\end{equation}
This is exactly a smooth binary neural response.  The nonlinearity is not imposed by a software-level choice of activation; it is the normalized interpolation between two vacua of the field theory.  The output $h\simeq0$ corresponds to one asymptotic vacuum sector of the readout, and $h\simeq1$ to the other.  The transition width is controlled by the kink mass scale and therefore by the curvature of the potential at the vacua.

A supervised binary target $y\in\{0,1\}$ may be represented by adding a data-induced energy
\begin{equation}
    E_{\rm data}(q)=\frac{\gamma}{2}\left[h(q)-y\right]^2,
\end{equation}
so that the collective coordinate relaxes according to
\begin{equation}
    \frac{dq}{ds}=-G^{-1}(q)\frac{d}{dq}\left[U_{\rm eff}(q)+E_{\rm data}(q)\right].
\end{equation}
The consequences are physically transparent.  Correctly classified data correspond to collective coordinates sitting in the appropriate basin of the deformed solitonic potential, while ambiguous data lie near the domain-wall transition region where $h\simeq1/2$.  The margin of the classifier is therefore controlled by the width and energetic stiffness of the kink.

\subsection{Example II: two interacting solitonic inputs and an induced two-variable map}

A second minimal construction uses two input solitons and one output readout.  Let $q_1$ and $q_2$ be two slow collective coordinates and let $q_o$ be the output coordinate.  Expanding the effective energy to lowest nontrivial order gives
\begin{equation}
    U_{\rm eff}(q_o,q_1,q_2)=U_o(q_o)-J_{o1}q_oq_1-J_{o2}q_oq_2-J_{o12}q_oq_1q_2+\cdots .
    \label{eq:two_input_energy}
\end{equation}
The stationarity condition for the output coordinate is
\begin{equation}
    \frac{\partial U_o}{\partial q_o}=J_{o1}q_1+J_{o2}q_2+J_{o12}q_1q_2+\cdots .
\end{equation}
If the output profile is kink-like, the projected output becomes
\begin{equation}
    h_o=\sigma\!\left(W_{o1}h_1+W_{o2}h_2+W_{o12}h_1h_2+b_o\right),
    \label{eq:two_input_readout}
\end{equation}
where the coefficients are determined by the corresponding interaction integrals.  Equation~\eqref{eq:two_input_readout} illustrates a key point: nonlinear solitonic interactions naturally generate not only ordinary affine couplings but also higher-order effective features.  A conventional neural network would introduce these features by architecture design or by depth; in the solitonic construction they arise from the nonlinear interaction energy.

\begin{figure}[t]
\centering
\includegraphics[width=0.60\textwidth]{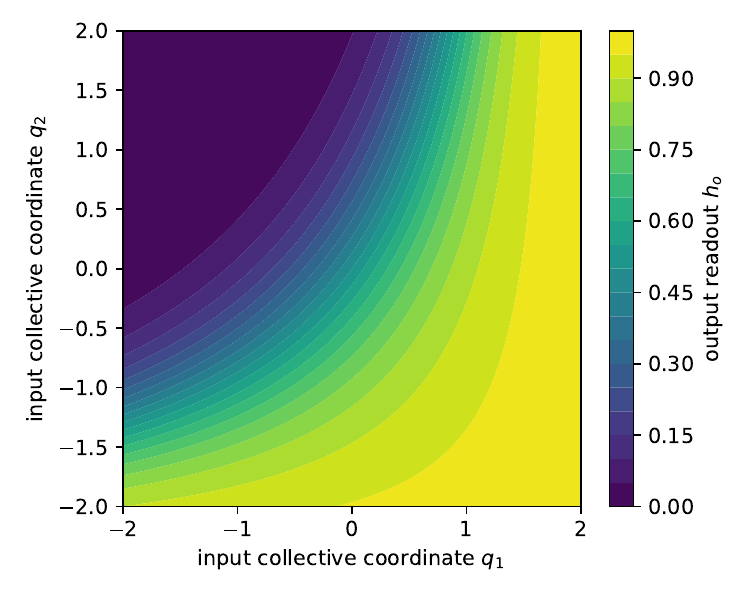}
\caption{Illustrative two-input solitonic readout.  The plotted surface represents a projected output $h_o=\sigma(2q_1-1.5q_2+0.8q_1q_2)$ generated by linear and bilinear interaction terms in the effective energy.  The figure is not a numerical simulation of a specific material platform; it visualizes how solitonic interaction coefficients become neural couplings and nonlinear features.}
\label{fig:two_soliton_readout}
\end{figure}

These examples also clarify the physical meaning of ``application'' in the present paper.  We are not applying an existing neural network to a dataset.  We are applying the solitonic construction itself to two minimal field-theoretic situations and extracting the corresponding neural maps.  The first example produces a single nonlinear neuron; the second produces a small interaction-induced network.  Larger feedforward architectures are obtained by repeating the same projection across several interacting solitonic sectors or across several discrete evolution steps.

\section{Beyond a single kink: coupled fields and richer neural units}\label{sec:coupled_fields}

\subsection{Two-field potentials and internal coordinates}

A single $\phi^4$ kink is sufficient to derive a $\tanh$ activation, but realistic solitonic media may contain several coupled fields.  Consider two scalar fields $\phi$ and $\chi$ with action
\begin{equation}
    S[\phi,\chi]=\int\dd^{d+1}x\left[\frac{1}{2}\partial_\mu\phi\partial^\mu\phi+\frac{1}{2}\partial_\mu\chi\partial^\mu\chi - V(\phi,\chi)\right].
    \label{eq:two_field_action}
\end{equation}
If $V(\phi,\chi)$ admits kink, lump, or oscillon-like solutions, the collective coordinates may include internal amplitudes and relative phases in addition to positions.  Coupled scalar models are well known to produce richer localized structures than single-field models, including long-lived oscillons and oscillating kinks with tunable shapes~\cite{CorreaDutra2016,CorreaDutraFredericoMalomed2019}.

For neural construction this matters because internal coordinates can play the role of gates.  A two-field solitonic unit may have an output
\begin{equation}
    h_i=\Pi_i[\Phi_i(x;q_i),X_i(x;q_i)],
    \label{eq:two_field_readout}
\end{equation}
where $\Phi_i$ and $X_i$ denote the two field components of the localized unit.  If one component controls the center or width while the other controls the readout amplitude, the resulting unit resembles a gated activation.  Thus coupled-field solitons naturally enlarge the class of neural nonlinearities beyond scalar $\tanh$ units.

\subsection{Oscillons as metastable computational units}

Oscillons are spatially localized, long-lived, time-dependent field configurations that arise in nonlinear scalar theories~\cite{Gleiser1994,CorreaDutra2016}.  They are not topological solitons in the strict kink sense, but they can possess lifetimes much longer than the microscopic oscillation period.  If the computation time is short compared with the oscillon lifetime, an oscillon can serve as a metastable neural unit.  The activation may then be defined by an envelope, a phase-averaged amplitude, or a stroboscopic readout,
\begin{equation}
    h_i(\tau)=\frac{1}{T}\int_\tau^{\tau+T}\dd t\int\dd x\,\rho_i(x)\phi_i(x,t).
    \label{eq:oscillon_readout}
\end{equation}
This possibility links the present framework to nonlinear wave computation.  Topological kinks provide robust monotone activations; oscillons provide dynamical resonant activations; nonlinear lattices provide spatially structured coupling graphs.  Each case is a different reduction of a nonlinear field theory to computational degrees of freedom.

\section{Phenomenological implications, limitations, and tests}\label{sec:phenomenology}

\subsection{Implications for architecture design}

The construction suggests a design principle: choose activation functions and coupling patterns from stable nonlinear field configurations rather than from purely algebraic convenience.  Kink profiles give monotone saturating activations; oscillon envelopes give localized oscillatory activations; compacton-like profiles give finite-support nonlinearities; nonlocal solitons give kernel-mediated response functions; and lattice solitons give naturally sparse or structured coupling graphs~\cite{KartashovMalomedTorner2011,CorreaDutraFredericoMalomed2019,Malomed2022}.  This does not imply that every such activation will outperform standard choices.  It implies that there is a physically organized search space for neural nonlinearities.

\subsection{Possible physical platforms}

The framework is compatible with several physical platforms where nonlinear localized waves and tunable couplings occur.  Nonlinear optical media support spatial and temporal solitons; Bose-Einstein condensates support matter-wave solitons; Josephson and magnetic systems can support domain walls and fluxons; mechanical and electrical nonlinear lattices can support discrete breathers and intrinsic localized modes~\cite{DauxoisPeyrard2006,KartashovMalomedTorner2011,Malomed2016}.  A solitonic neural device would not store weights as abstract floating-point values.  It would encode them as controllable interaction strengths, separations, refractive-index profiles, external fields, or boundary conditions.

\subsection{Quantitative checks and nontrivial predictions}

The proposal becomes testable when a field model and projection are fixed.  It predicts: (i) the activation shape should match the solitonic profile or scattering response; (ii) measured interaction coefficients should predict the effective weights; (iii) perturbations that preserve the solitonic sector should leave the computational output more stable than perturbations that destabilize the sector; and (iv) changing the nonlinear potential should change the activation family in a calculable way.  These statements are stronger than a verbal analogy because they can fail.

\subsection{Limitations}

The present construction is not a universal derivation of all deep learning.  It applies to architectures that can be represented as reduced dynamics of localized nonlinear modes.  It also relies on truncating fluctuations, assuming a controlled collective-coordinate approximation, and using a leading-order expansion of the interaction energy.  Strong radiation, soliton annihilation, chaotic scattering, or large fluctuation backreaction can invalidate the finite-dimensional neural description.  These limitations are important because they define the regime in which the proposed field-theoretic origin is meaningful.

\section{Conclusion}\label{sec:conclusion}

We have developed a field-theoretic construction in which a class of artificial neural networks arises from solitonic degrees of freedom in nonlinear scalar field theory. The central point of the paper is not that neural networks admit a suggestive description in the language of solitons. Rather, the result is that the elementary ingredients of a neural architecture can be obtained by a controlled reduction of a nonlinear continuum theory to a finite-dimensional solitonic sector. Starting from an action functional, one restricts the configuration space to localized stable solutions, introduces collective coordinates, computes the induced moduli-space metric and effective interaction energy, and then projects the reduced dynamics onto computational variables. The neural layer appears only at the end of this sequence, as the effective input-output map of interacting solitonic modes.

This construction changes the status of the usual neural ingredients. The neuron is not postulated as an abstract scalar unit, but emerges as a projected observable of a localized field configuration. The activation function is not chosen phenomenologically, but is inherited from the nonlinear response profile or scattering map of a soliton. The weight matrix is not introduced as an arbitrary array of parameters, but arises from overlap integrals, Hessians, or interaction coefficients in the reduced solitonic energy. The bias is naturally associated with external sources, boundary forcing, vacuum asymmetry, or background deformations. Finally, the depth of a feedforward architecture is interpreted as a discrete evolution parameter on the solitonic moduli space.

The \(\phi^4\) kink provides the minimal explicit realization of this program. Its normalized profile gives the \(\tanh\) activation, while its shifted and normalized profile gives the logistic sigmoid. More importantly, the kink example shows how a nonlinear field configuration can be converted into a computational unit through a projection map \(h=\Pi[\Phi]\). When several localized modes interact, the reduced interaction energy induces effective couplings among the projected variables. Under a discrete interaction-response or operator-splitting approximation to collective-coordinate gradient flow, these projected variables obey a map of the feedforward form
\[
h^{(\ell+1)}
=
\sigma
\left(
W^{(\ell)}h^{(\ell)}+b^{(\ell)}
\right).
\]
Thus, the familiar affine-plus-nonlinearity structure of a neural layer is recovered as an effective consequence of solitonic dynamics, rather than assumed as a primitive computational rule.

The conceptual significance of the construction is that it provides a physical route from nonlinear localized excitations to neural-network structure. In this view, artificial neural networks are not merely formal compositions of scalar nonlinearities; at least for the class considered here, they can be understood as effective discrete dynamics on a reduced space of solitonic configurations. This interpretation also gives a natural physical meaning to robustness. Since solitons owe their persistence to energetic barriers, nonlinear coherence, and, in topological cases, boundary conditions or conserved charges, the associated projected computational states can inherit a degree of stability against microscopic perturbations. Robustness of the neural map is then tied to stability properties of the underlying solitonic sector.

The construction also clarifies its own limitations. It does not imply that every neural network is uniquely generated by a solitonic field theory, nor that every activation function has a topological origin. The claim is more precise: whenever a nonlinear field theory admits localized stable solutions and a separation of scales allowing a collective-coordinate reduction, one obtains a natural mechanism for generating neural units, activation profiles, interaction weights, and layer-wise maps. The correspondence is therefore constructive, not merely linguistic, and model-dependent in a controlled way.

Several directions follow from this framework. On the theoretical side, one may extend the construction to multi-component fields, non-Abelian solitons, Q-balls, vortices, oscillons, domain-wall networks, and nonlinear lattices, where richer internal moduli may generate multi-channel or attention-like architectures. On the computational side, one may design solitonic activation functions and coupling patterns directly from effective field potentials, and compare their trainability, expressivity, and robustness with standard neural architectures. On the physical side, the framework suggests possible implementations in nonlinear optical systems, condensed-matter media, metamaterials, and lattice platforms where localized nonlinear excitations can be created, coupled, and measured.

The main conclusion is therefore that a neural network can be viewed, in a precise and constructive sense, as the finite-dimensional effective dynamics of interacting solitonic configurations. The bridge from field theory to computation is provided by collective-coordinate reduction and projection. This offers a field-theoretic origin for a class of neural architectures and opens a path toward neural models whose units, nonlinearities, and couplings are derived from the physics of localized nonlinear structures.

\begin{acknowledgments}

The author gratefully acknowledges financial and institutional support from Capgemini. The author also thanks Capgemini for supporting research activities at the interface of theoretical physics, quantum technologies, and advanced scientific computing. The views expressed in this work are those of the author and do not necessarily represent the official position of Capgemini.

\end{acknowledgments}

\appendix

\section{Detailed derivation of the kink energy}\label{app:kink_energy}

For the $\phi^4$ kink,
\begin{equation}
    \phi_K=v\tanh\left(\frac{m}{2}(x-X)\right),
    \qquad m=\sqrt{2\lambda}v.
\end{equation}
The derivative is
\begin{equation}
    \phi_K'=\frac{mv}{2}\sech^2\left(\frac{m}{2}(x-X)\right).
\end{equation}
Using the BPS equation, the energy density is $(\phi_K')^2$, so
\begin{align}
    E_K&=\int_{-\infty}^{\infty}\dd x\,(\phi_K')^2 \\
    &=\frac{m^2v^2}{4}\int_{-\infty}^{\infty}\dd x\,\sech^4\left(\frac{m}{2}(x-X)\right).
\end{align}
Let $u=m(x-X)/2$, so $\dd x=2\dd u/m$.  Then
\begin{equation}
    E_K=\frac{mv^2}{2}\int_{-\infty}^{\infty}\dd u\,\sech^4u.
\end{equation}
Since $\int_{-\infty}^{\infty}\sech^4u\,\dd u=4/3$,
\begin{equation}
    E_K=\frac{2mv^2}{3}=\frac{2\sqrt{2\lambda}v^3}{3}.
\end{equation}
This finite energy is the mass of the kink in the classical theory.

\section{From Hessian interactions to neural weights}\label{app:hessian_weights}

Let the reduced energy near a reference configuration be
\begin{equation}
    U(q)=U_\star+F_a\delta q^a+\frac{1}{2}K_{ab}\delta q^a\delta q^b+\cdots.
\end{equation}
Split the coordinates into response coordinates $z_i$ and readout coordinates $h_i=\sigma_i(z_i)$.  The quasi-static equation for $z_i$ is
\begin{equation}
    0=\frac{\partial U}{\partial z_i}
    =\mu_i z_i+\sum_j\tilde K_{ij}h_j-J_i+\cO(h^2,z^2).
\end{equation}
Solving to leading order gives
\begin{equation}
    z_i=\sum_jW_{ij}h_j+b_i+\cO(h^2),
    \qquad
    W_{ij}=-\frac{\tilde K_{ij}}{\mu_i},
    \qquad
    b_i=\frac{J_i}{\mu_i}.
\end{equation}
Substitution into $h_i=\sigma_i(z_i)$ gives the neural unit.  Higher-order corrections generate polynomial or tensor-valued interactions.

\section{Bibliographic note on scope}\label{app:bib_scope}

The bibliography is intentionally ordered by first appearance and broad enough to locate the construction in three communities: classical soliton theory, nonlinear localized waves, and mathematical theory of neural networks.  The soliton references establish the legitimacy of finite-energy localized sectors and collective coordinates.  The nonlinear-lattice and oscillon references motivate tunable localized modes beyond the static kink.  The neural-network references clarify how the present construction differs from existing continuum-limit and field-theoretic descriptions of already-defined networks.

\end{document}